\documentstyle{elsart}
 
\def\Journal#1#2#3#4{{#1} {\bf #2}, #3 (#4)}

\def\APP{{\em Acta Phys. Pol.} B}
\def\NCA{\em Nuovo Cimento}
\def\NIM{\em Nucl. Instrum. Methods}
\def\NIMA{{\em Nucl. Instrum. Methods} A}
\def\NPA{{\em Nucl. Phys.} A}
\def\NPB{{\em Nucl. Phys.} B}
\def\PLB{{\em Phys. Lett.}  B}
\def\PRL{\em Phys. Rev. Lett.}
\def\PR{\em Phys. Rev.} 
\def\PREP{\em Phys. Rep.} 
\def\PRC{{\em Phys. Rev.} C}
\def\PRD{{\em Phys. Rev.} D}
\def\ZPC{{\em Z. Phys.} C}

\newcommand{\checkthis}{\marginpar{$\Leftarrow$\em Check}}

\begin{document}
\hspace{9.8 cm}FZJ--IKP(TH)--1998--20
\begin{frontmatter}
 
\title{On the treatment of $NN$ interaction effects in
meson production in $NN$ collisions}
 
\author{C. Hanhart$^{a,b,\dagger}$ 
and K. Nakayama$^{b,c,\dagger \dagger}$}

{\small $^a$Institut f\"ur Theoretische Kernphysik, Universit\"at Bonn,}\\
{\small  D--53115 Bonn, Germany} \\
{\small $^b$Institut f\"{u}r Kernphysik, Forschungszentrum J\"{u}lich GmbH,}\\ 
{\small D--52425 J\"{u}lich, Germany} \\
{\small $^c$Department of Physics and Astronomy, University of Georgia,}\\
{\small  Athens, GA 30602, USA} \\
{\small $\dagger$ e--mail: \tt c.hanhart@fz-juelich.de} \\
{\small $\dagger \dagger$ e--mail: \tt nakayama@archa11.cc.uga.edu} \\

\begin{abstract}
We clarify under what circumstances the nucleon--nucleon final state interaction 
fixes the energy dependence of the total cross--section for the reaction $NN \to 
NNx$ close to production threshold, where $x$ can be any meson whose interaction
with the nucleon is not too strong. We strongly criticize the procedure used 
recently by several authors to include the final state interaction in the reactions 
under discussion. In addition, we give a formula that allows one to estimate the 
effect of the initial state interaction for the production of heavy mesons.
\end{abstract}

\end{frontmatter}

{\it Keywords: final state interaction, initial state interaction, meson production}
 
PACS: 25.40.-h, 13.75.Cs

As early as 1952 K. Watson pointed out under what circumstances one expects the
final state interaction to strongly modify the energy dependence of the total
production cross--section $NN \to NNx$ \cite{watson} given by
$$
\sigma_{NN \to NNx}(\eta) \propto \int_0^{m_x\eta}d\rho (q')|A(E,p')|^2 \ .
$$
In the above equation $q'$ is the momentum of the outgoing meson and $A(E,p')$ is 
the $NN \to NNx$ transition amplitude, which, for future convenience, is expressed
as a function of the total energy $E$ and the relative momentum of the two nucleons
in the final state $p'$ (note, that $p'$ and $q'$ are related to each other via 
energy conservation). The phase space is denoted by $d\rho(q')$ and $\eta$ denotes 
the maximum momentum of the emitted meson in units of its mass. Watson \cite{watson} 
argues that if there is a strong and attractive force between two of the outgoing 
particles, as is the case for the reactions under consideration, the energy 
dependence of the total cross--section is determined by the phase space and the 
energy dependence of the relevant attractive interaction, i.e.,
\begin{equation}
\nonumber
\sigma_{NN \to NNx}(\eta) \propto \int_0^{m_x\eta}d\rho (q')|T(p',p')|^2  
\propto \int_0^{m_x\eta}d\rho (q')\left(\frac{\sin{\delta(p')}}{p'}\right)^2 \ ,
\label{sendep}
\end{equation}
In the above equation $T(p',p')$ is the on--shell $NN$ $T$--matrix, $\delta (p')$ 
denotes the $NN$ phase shifts at the energy, $2E(p')$, of the final $NN$ subsystem 
(here restricted to s--waves), where  $E(p') \equiv {p'^2\over 2 m}$, with $m$ 
denoting the nucleon mass. When data for the reaction $pp \to pp\pi^0$ close to 
threshold became available \cite{IU1} eq. \ref{sendep} indeed turned out to give the
correct energy dependence of the total cross--section \cite{MuS}. Several authors 
\cite{moalem1,mosel,sibir1,sibir2,moalem2,ulfneu}
concluded from this observation that it is appropriate to calculate the transition
$NN \to NNx$ to lowest order in perturbation theory and just include the final state
interaction (FSI) by using a formula of the type in eq. \ref{sendep}; they implement
the FSI by use of just the on--shell $NN$ $T$--matrix, not only to get the right 
energy dependence of the cross--section, but also to get the strength of the matrix 
elements. In this letter we criticize this procedure. We shall demonstrate that the 
observation that the energy dependence of the cross--section is given by the {\it 
on--shell} FSI does not necessarily imply that the strength of the matrix elements 
is also determined by the on--shell $NN$ interaction. We also show that Watson's 
requirement that the FSI be attractive in order to obtain the energy dependence of 
the cross--section given by eq. \ref{sendep} is unnecessary. Finally we give an 
expression that allows one to estimate the effect of the initial state interaction 
(ISI) on the reaction $NN \to NNx$, with $x$ any meson heavier than the pion, in 
terms of the (on--shell) $NN$ scattering phase shifts and inelasticities.

\begin{figure}[h]
\vspace{5cm}
\includegraphics{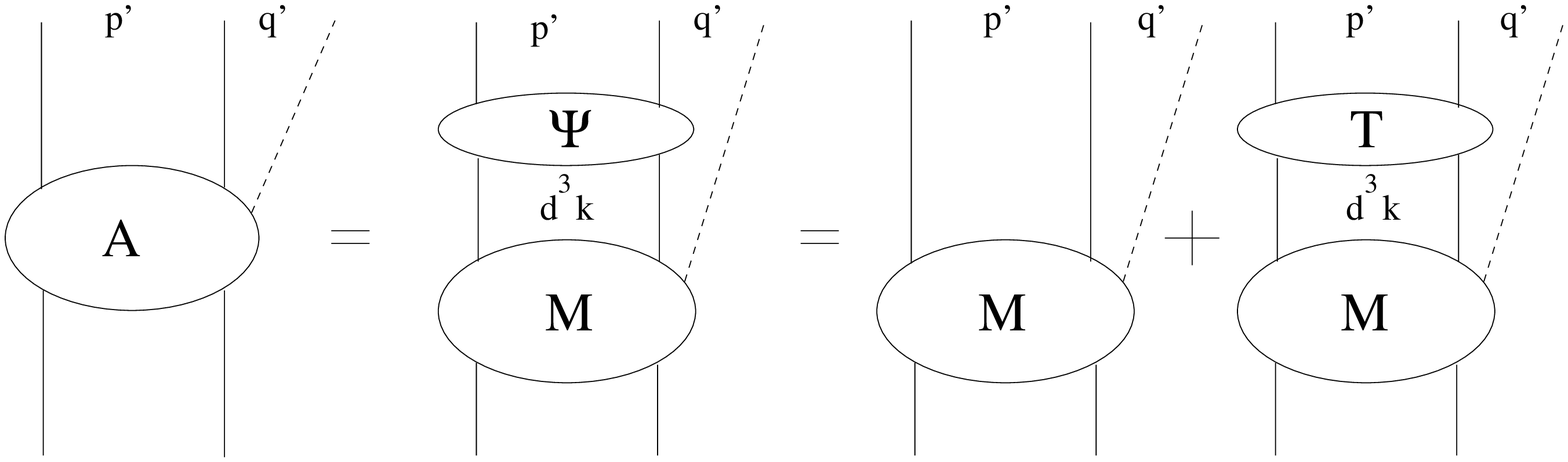}
\caption{\it{Decomposition of the production amplitude in the final state NN interaction
and the production part. Here $\Psi$ denotes the nucleon--nucleon wave function and
$T$ stands for the $NN$ $T$--matrix.}}
\label{decomp}
\end{figure}

The starting point of the present investigation is the decomposition of the total 
transition amplitude into a production  amplitude, hereafter called $M$, and the 
$NN$ FSI (see also Fig. \ref{decomp}). As $M$ is not specified, no approximation is 
involved in this decomposition. Schematically we can write
$$
A \ = \ M \ + \ TGM \ .
$$
An integration over the intermediate momenta is needed to evaluate the second term 
on the right hand side. This is actually the term where the off--shell information 
of both the $NN$ $T$--matrix and $M$ enters, as will become clear below. To be 
concrete,  we use non--relativistic kinematics for simplicity. The generalization
to a fully relativistic treatment is straightforward and does not provide any
new insights. In addition, since we only want to investigate effects
of the FSI on  the energy dependence of the total cross--section, overall constant 
factors are dropped. Using \cite{gold}
$$
G(E,k) = {\bf P}\frac{1}{E \ - \ 2E(k)} \ - \ i \ \pi \delta (E-2E(k)) \ ,
$$
where {\bf P} denotes the principal value, we write the total transition amplitude 
$A$ in the form
\begin{equation}
A(E,p') \ = \ M(E,p')\left\{ 1 - i\kappa (p') T(p',p')[1 + \frac{i}{ap'}
{\cal P}(E,p')] \right\}  \ .
\label{intermediate}
\end{equation}
where $\kappa (p') =\frac{\pi p' m}{2}$ is the phase space density; the factor of 
$1/ap'$, with $a$ denoting the low-energy $NN$ scattering length, has been introduced
for future convenience. Also, for convenience, we display only those arguments of $M$ 
that are relevant for the present discussion, that is the total energy $E$ and the 
relative momentum $p'$ of the two nucleons in the final state. As pointed out 
in ref. \cite{watson}, $M$ depends weakly on $E$ if the production mechanism is short 
ranged. In the above equation, all the off--shell effects are contained in the function
${\cal P}(E,p')$, whose explicit form is
\begin{equation}
{\cal P}(E,p')  =  \frac{ap'}{\kappa (p')}{\bf P}\int_0^\infty \ dk \frac{k^2f(E,k)}{E-2E(k)} 
=   \frac{2a}{\pi}\int_0^\infty \ dk 
\left[\frac{k^2 f(E,k)-p'^2}{p'^2-k^2}\right] \ ,
\label{defP}
\end{equation}
with the function $f$  defined as
\begin{equation}
f(E,k) \ = \ \frac{T(p',k)}{T(p',p')}\frac{M(E,k)}{M(E,p')} \ 
= \  \frac{K(p',k)}{K(p',p')}\frac{M(E,k)}{M(E,p')} \ .
\label{deff}
\end{equation}
The last equality in the above equation follows from the half--off--shell unitarity
relation of the $NN$ $T$--matrix, namely
$$
T(p',k) = \frac{1}{2} \left( \eta (p') e^{2i\delta(p')}+1\right) K(p',k)
$$
with  the $K$--matrix real by definition. Therefore, all of the imaginary part of $f$
-- and therefore of $\cal P$ -- is introduced by the production amplitude $M$. In 
the latter formula use has been made of the fact that the on--shell $T$--matrix and 
the phase shifts are related by
\begin{equation}
\kappa(p') T(p',p') \ =  \ \frac{i}{2}\left( \eta (p') e^{2i\delta(p')}-1\right) \ ,
\label{phsft}
\end{equation}
where $\eta (p')$ denotes the inelasticity.

Substituting eq. \ref{phsft} into eq. \ref{intermediate}, we get for the transition 
amplitude
\begin{eqnarray}
\nonumber
 A(&E&, p')  =  \frac{1}{2}M(E,p') e^{i\delta (p')}  \\ & \times & 
\left[ \left( \eta (p') e^{i\delta(p')}+e^{-i\delta(p')}\right) -
\frac{1}{i} \left( \eta (p') e^{i\delta(p')}-e^{-i\delta(p')}\right) 
\frac{1}{ap'}{\cal P}(E,p')\right] \ .
\label{aofpweta}
\end{eqnarray}

For energies near the production threshold energy one has $\eta (p') = 1$, so 
that,
\begin{equation}
A(E,p') \ = \  M(E,p') e^{i\delta (p')} \left[ \cos (\delta (p')) - 
             \frac{\sin (\delta (p'))}{ap'}{\cal P} (E,p')\right] \ .
\label{aofp}
\end{equation}
Using the effective range expansion 
\begin{equation}
p'\cot (\delta (p')) = -{1\over a} + {1\over 2}\Lambda^2 \sum_{n=0}^\infty
                        r_n\left({p'^2\over\Lambda^2}\right)^{n+1}
\label{erexp}
\end{equation}
eq. \ref{aofp} can be further reduced to
\begin{eqnarray}
\nonumber
A(E,p') & = & -  M(E,p') e^{i\delta (p')} \left({\sin (\delta (p')) \over
ap'}\right)\\ & \phantom{=} & \hspace{4cm} \times
\left[ \ {\cal P} (E,p') + 1 - {1\over 2}a r_o p'^2 - ... \ \right] \ .
\label{aofper}
\end{eqnarray}

This is the central formula of the present discussion. It reveals a number of 
important features. First of all, it shows that the energy dependence of the 
total cross--section is, indeed, given by eq. \ref{sendep} as has been shown 
by Watson, provided the production amplitude $M(E,p')$ and the function 
${\cal P}(E,p')$ have a weak energy dependence compared to that due to the FSI.
Secondly, it is not necessary that the FSI be attractive in order for the 
total cross section to have the energy dependence given by eq. \ref{sendep}: 
as long as $M(E,p')$ and ${\cal P}(E,p')$ have a weak energy dependence, the 
energy dependence of the total cross--section will be given by the FSI times 
phase space for $p'^2 \ll (ar_0)^{-1}$. Thirdly, and most relevant to the present 
discussion, the above formula also shows that the strength of the amplitude 
$A(E,p')$ depends on the function ${\cal P}(E,p')$.  Using just the on--shell 
$T$--matrix as the FSI instead of the full half--off--shell $T$--matrix means 
setting the function ${\cal P}(E,p')$ to zero. As has been mentioned before, the 
function ${\cal P}(E,p')$ summarizes all the off--shell effects of the FSI and
production amplitude. As such, it is an unmeasurable and model--dependent quantity. 
In particular, it depends on the particular regularization scheme used. For example,
in conventional calculations based on meson--exchange models, where the regularization
is done by introducing  form factors, the function ${\cal P}(E,p')$ is very 
large and cannot be neglected\footnote{We checked this numerically.}.
 Other regularization schemes, however, may yield a 
vanishing function ${\cal P}(E,p')$. Since the total amplitude $A(E,p')$ should not 
depend on the particular regularization scheme, the production amplitude $M(E,p')$ 
in eq. \ref{aofper} must depend on the regularization scheme in such a way to 
compensate for the regularization dependence of ${\cal P}(E,p')$. At this stage one 
has to conclude that the procedure of just evaluating $M$ in the on--shell tree 
level approximation and simply multiplying it with the on--shell $NN$ $T$--matrix 
without consistency between the $NN$ scattering and production amplitudes (as it 
was done in \cite{moalem1,mosel,sibir1,sibir2,moalem2,ulfneu}) is not acceptable in 
order to obtain {\it quantitative predictions}. In the appendix  we develop a  
simple model in order to gain more insight on this issue.

All the above considerations are not restricted to the $NN$ final states; whenever 
there is a strong two--particle correlation in the final state,  the energy 
dependence of a total production cross--section is given by the on--shell phase 
shifts of two of the outgoing particles. This condition is 
for example also met in the 
reaction $pp \to pK\Lambda$, as demonstrated in ref. \cite{pklam}.

The situation is very different for the effect of the $NN$ interaction, responsible
for the initial state distortions. Since the kinetic energy of the initial state has
to be large enough to produce a meson, the $NN$ ISI is evaluated at 
large energies. Therefore, in this regime we expect the variation with energy of the
ISI to be small.
\footnote{In case of pion production the phase shifts of the $^3P_0$ 
partial wave, which is the initial state for the s--wave $\pi^0$ production, still 
vary reasonably rapidly with energy. Therefore we do not expect the principal value 
integral to be small.} 
At least in the case of meson--exchange models this implies a flat off--shell 
behavior of the $NN$ $T$--matrix at a given energy, in which case the principal 
value integral is expected to be small, as can be seen from eq. \ref{defP}. It is 
this observation that allows us to use eq. \ref{aofpweta} to estimate the effect of
the ISI on the total production cross--section for the production of heavier mesons.
The ISI therefore leads to a reduction of the total cross--section of the order of
\begin{eqnarray}
\nonumber
\lambda & = & \left| \frac{1}{2} e^{i\delta_L (p)} 
              \left( \eta_L (p) e^{i\delta_L(p)}+e^{-i\delta_L(p)}\right) \right|^2
 \\     & = & \eta_{L}(p) \cos ^2(\delta_{L}(p)) + 
              \frac{1}{4}[1-\eta_{L}(p)]^2 \leq \frac{1}{4}[1+\eta_{L}(p)]^2 \ ,
\label{isieff}
\end{eqnarray}
where $p$ denotes the relative momentum of the two nucleons in the initial state 
with the total energy $E$. The index $L$ indicates the quantum numbers of the 
corresponding initial state. Note that, for production reactions close to threshold, 
selection rules strongly restrict the number of allowed initial states. In the 
literature there is one example that quantifies the effect of the ISI for meson 
production reactions, namely ref. \cite{lee}, where the reaction $pp \to pp\eta$ is 
studied. The inclusion of the ISI in this work leads to a reduction of the total 
cross--section by roughly a factor of 0.3.  At threshold only the $L=\ ^3P_0$ state 
contributes to the ISI. The phase shifts and inelasticities given by the model used 
in \cite{lee} for the ISI are $\delta_L(p) = -60.7^o$ and $\eta_L(p) = 0.57$ 
\cite{lee2} at $T_{Lab} = 1250$ MeV. These values agree with the phase shift 
analysis given by the SAID program \cite{said}. Both phase shifts and the 
inelasticity vary by 10\% only over an energy range of $500$ MeV \cite{said}. Using 
the above mentioned values for $\delta_L(p)$ and $\eta_L(p)$ we get for the 
reduction factor $\lambda$, defined in eq. \ref{isieff}, a value of 0.2. Therefore, in
the case of the kinematics of the ISI for the $\eta$ production, the principal value
integral -- within the meson--exchange model used -- indeed turns out to be a 
correction of the order of 20\% compared to the leading on--shell contribution. 

In summary, we have demonstrated that the way the FSI is treated in a series of 
recent papers is unjustified to achieve quantitative predictions. It should be clear,
however, that this does not fully disregard the results of 
refs. \cite{moalem1,mosel,sibir1,sibir2,moalem2,ulfneu}: For the purpose of 
investigating the relative importance of different contributions to the production 
process the approach used there is still justified.  Our criticism is not to the 
result of ref. \cite{watson}. In fact, our function ${\cal P}(E,p')$ appearing in 
eq. \ref{intermediate} is related to the factor $(f(r),\bar R)$ in eq. (32) of ref. 
\cite{watson}, where $f(r)$ accounts for the short--range behavior of the strongly 
interacting particles in the final state. Note that Watson \cite{watson} does not 
give a prescription how to calculate the overlap integral $(f(r),\bar R)$, which 
would be required to fix the overall normalization. We emphasize that we do $not$ 
claim that off--shell effects are measurable \cite{fear}. The result of this paper 
is the demonstration of the necessity to properly account for loop effects of the 
FSI in situations where the latter strongly influences the energy dependence of the 
total cross--section as in meson production in $NN$ collisions. In addition, we have
given a compact formula that allows one to estimate the effect of the ISI in terms 
of the phase shifts and inelasticities of $NN$ scattering. This formula should prove
to be useful for theoretical investigations of the production of heavy mesons close 
to their production threshold.

{\bf Appendix}

In this appendix we will use a simple model to demonstrate the need for a consistent
treatment of both the $NN$ scattering and production amplitudes in order to obtain 
quantitative predictions of meson--production reactions.

Let us assume a separable $NN$ potential
\begin{equation}
V(p', k) = \alpha g(p') g(k) \ ,
\label{sepV}
\end{equation}
where $\alpha$ is the coupling constant and $g(p')$ an arbitrary real function of 
$p'$. With this potential the $T$--matrix scattering equation can be readily solved 
to yield
\begin{equation}
T(p', k) = {V(p', k)\over 1 - R(p') + i\kappa (p')V(p', p')} \ ,
\label{sepT0}
\end{equation}
with
\begin{equation}
R(p') \equiv  m{\bf P} \int_0^\infty \ dk' \ {k'^2 V(k',k')
                                                 \over p'^2 - k'^2} \ .
\label{Rint}
\end{equation}

Note that for an arbitrary
function $g(k)$, such as $g(k) \equiv 1$ as discussed 
below, $R(p')$ may become a divergent integral. Also, if the potential is 
derived from field theory in general, the principal value integral turns out to be
divergent. In these cases $R$ is to be understood as properly regularized.
The principal value integral $R(p')$ given above is therefore not only a model-dependent quantity, 
but also depends on the regularization scheme used.
 The condition that the 
on--shell $NN$ scattering amplitude should satisfy eq. \ref{phsft} relates this 
model-- and regularization--dependent quantity to the on--shell potential, 
$V(p', p')$, 
\begin{equation}
R(p') = 1 + \kappa(p') \cot(\delta(p')) V(p', p') \ ,
\label{RV}
\end{equation}
where it is assumed that $\eta(p')=1$. This shows that, for a given potential, the
regularization should be such that eq. \ref{RV} be satisfied in order to reproduce 
eq. \ref{phsft}. Indeed, in conventional calculations based on meson--exchange models,
where one introduces form factors to regularize the principal value integral, the 
cutoff parameters in these form factors are adjusted to reproduce the $NN$ 
scattering phase shifts through eq. \ref{phsft}. Conversely, for a given 
regularization scheme, the $NN$ potential should be adjusted such as to obey 
eq. \ref{RV}. This is the procedure used in effective field theories \cite{KSW}, 
where the coupling constants in the $NN$ potential become regularization--dependent.

We also assume that the production amplitude $M$ is given by a separable form
\begin{equation}
M(E, k)  =  \beta g(k) h(p) \ ,
\label{sepM} 
\end{equation}
where $\beta$ is the coupling constant and $h(p)$ an arbitrary function of the 
relative momentum $p$ of the two nucleons in the initial state. With
the $NN$ $T$--matrix and production amplitude given above the function ${\cal P}(E,p')$
 given by eq. \ref{defP} can be written as
\begin{equation}
{\cal P}(E,p')  =  ap'\left( {R(p')\over R(p') - 1} \right) \cot(\delta(p')) \ .
\label{seP}
\end{equation}
Thus, since $R(p')$ is a regularization--dependent quantity, so is the function 
${\cal P}(E, p')$, as it was argued in the main section. Inserting this
 into eq. \ref{aofp}, and using eq. \ref{RV}, we can express the total transition 
amplitude as 
\begin{equation}
A(E,p') = - {1\over \kappa (p')} e^{i\delta(p')}\sin(\delta(p'))
          \left({M(E,p')\over V(p',p')}\right) \ .
\label{sepA}
\end{equation}

Eq. \ref{sepA} is the desired formula for our discussion. It allows us to study the 
relationship between the $NN$ potential and the production amplitude $M(E,p')$ 
explicitly as different regularization schemes are used. For this purpose let us 
study the simplest case of a contact $NN$ potential (setting the function $g=1$ in 
eq. \ref{sepV})  in the limit of $p' \rightarrow 0$. If we regularize the 
integrals by means of the power divergent subtraction (PDS) scheme \cite{KSW} we get
$$
R = - \frac{a\mu}{1-a\mu} \ ,
$$
where $\mu$ denotes the regularization scale. Substituting this result into 
eq. \ref{RV}, we obtain  
$$
\alpha =\left( {2a \over \pi m} \right) {1 \over 1 - a\mu}
$$
for the $NN$ coupling strength. Note that for $\mu = 0$ the PDS scheme
reduces to that of minimal subtraction \cite{KSW}. Since the total production 
amplitude $A$ should not depend on the regularization scale we immediately read off 
from eq. \ref{sepA} that
$$
\beta \propto (1-a\mu )^{-1} \ .
$$
Therefore the model clearly exhibits the point made in the main section, namely, the 
requirement to calculate both the production amplitude and the FSI consistently in 
order to make quantitative predictions. 
                                  
Eq. \ref{seP} allows to calculate explicitly the function ${\cal P}(E,p')$. We have
$$
{\cal P} = -a\mu \ .
$$
The minimal subtraction scheme, therefore, leads to ${\cal P} = 0$. The authors of 
ref. \cite{KSW} argue, however, that $\mu \simeq m_\pi$ in order to provide a proper 
counting scheme for the $NN$ interaction. In this case, however, we have 
${\cal P} \simeq 6$ (here we used the pp scattering length of $a=-7.8$ fm).

{\bf Acknowledgments}

We thank J. Durso, J. Haidenbauer, Th. Hemmert and N. Kaiser for useful discussions
and W. Melnitchouk for careful reading of the manuscript.
One of the authors (C.H.) is grateful for the financial
 support by COSY FFE--Project Nr. 41324880.

%
%
%

%
 
\end{document}